\documentclass[11pt]{article}
\usepackage{amsfonts}
\usepackage{amssymb}
\def\bdt{\dot \beta}
\def\adt{\dot \alpha}

\newfont{\bbbold}{msbm10 scaled \magstep1}

\def\bbC{\mbox{\bbbold C}}

\newfont{\goth}{eufm10 scaled \magstep1}

\def\a{\alpha}

\def\d{\delta}\def\D{\Delta}
\def\e{\epsilon}
\def\vf{\varphi}
\def\h{\eta}\def\z{\zeta}

\def\l{\lambda}

\def\p{\pi}

\def\t{\tau}
\def\th{\theta}

\def\be{\begin{equation}}\def\ee{\end{equation}}
\def\bea{\begin{eqnarray}}\def\eea{\end{eqnarray}}
\def\ba{\begin{array}}\def\ea{\end{array}}

\def\del{\partial}

\def\nno{\nonumber}
\def\del{\partial}
\let\la=\label

\def\bd{\begin{document}}
\def\ed{\end{document}}
\def\bea{\begin{eqnarray}}
\def\eea{\end{eqnarray}}
\def\ba{\begin{array}}
\def\ea{\end{array}}
\def\ft#1#2{{\textstyle{{\scriptstyle #1}\over {\scriptstyle #2}}}}
\def\fft#1#2{{#1 \over #2}}
\newcommand{\eq}[1]{(\ref{#1})}
\def\eqs#1#2{(\ref{#1}-\ref{#2})}
\def\det{{\rm det\,}}
\def\tr{{\rm tr}}
\newcommand{\ho}[1]{$\, ^{#1}$}
\newcommand{\hoch}[1]{$\, ^{#1}$}
\newcommand{\tamphys}{\it\small Center for Theoretical Physics,
Texas A\&M University, College Station, TX 77843, USA} 
\newcommand{\newton}{\it\small Isaac Newton Institute for Mathematical
Sciences, Cambridge, UK} 
\newcommand{\kings}{\it\small Department of Mathematics, King's College,
London, UK} 
\newcommand{\lapp}{\it\small LAPP, Annecy, France}
\newcommand{\cern}{\it\small Theory Division, CERN, Geneva, Switzerland}
\begin{document}
 \thispagestyle{empty}
 
\hfill{CERN-TH-2000-113} 

 \hfill{KCL-MTH-00-19}

 \hfill{LAPTH-791/2000}

 \hfill{\today}
\vspace{20pt} 

 \begin{center}
{\Large{\bf Extremal and next-to-extremal $n$-point correlators in 
four-dimensional SCFT}} \vspace{30pt}\\ {\large B. Eden\hoch{b}, 
P.S. Howe\hoch{a}, E. Sokatchev\hoch{b} and P.C. West\hoch{a,c}} 
\vspace{15pt} 
\begin{itemize}
 \item [$^a$] \kings
 \item [$^b$] {\it\small Laboratoire d'Annecy-le-Vieux de Physique
 Th{\'e}orique\footnote{UMR 5108 associ{\'e}e {\`a}
 l'Universit{\'e} de Savoie} LAPTH, Chemin de Bellevue, B.P. 110,
F-74941 Annecy-le-Vieux, France}
\item [$^c$]{\it\small Theory Division, CERN, Geneva, Switzerland} 
 \end{itemize}
\vspace{60pt} 
 {\bf Abstract}
\end{center}
It is shown that certain extremal and next-to-extremal $n$-point correlators in four-dimensional 
$N=2$ superconformal field theories are free. These results hold 
for any gauge group. 

{\vfill\leftline{}\vfill 
\pagebreak \setcounter{page}{1} 

\section{Introduction}

In the harmonic superspace approach to the study of correlation 
functions of gauge-invariant operators in four-dimensional 
superconformal field theories (with $N=2$ or $N=4$), the operators 
that can be investigated most easily are either products of 
hypermultiplets ($N=2$) or products of the $N=4$ field strength 
superfield $W$. It was originally conjectured in \cite{hw1} that 
the constraints of analyticity in the internal space 
(H-analyticity), generalised chirality (G-analyticity) and 
superconformal symmetry might be strong enough to determine 
correlation functions of such operators, at least for sufficiently 
low charges (the charge being proportional to the number of fields 
in the product). Although this has turned out not to be the case 
for generic values of the charges (see \cite{toappear} for a 
detailed discussion of $N=2$ four-point functions in this 
approach), it is true for certain special values of the charges. 

Correlation functions for these special values of the charges, 
called extremal correlators, were discussed in AdS supergravity in 
\cite{dfmmrext} and \cite{arufrov}, and the results found there, 
namely that these correlators are products of free propagators, 
were verified at one-loop order in perturbative field theory in 
\cite{biakov} and later to all orders, for four points, using 
harmonic superspace methods \cite{ourExt}. It was further shown in 
\cite{ourExt} that these simple results also hold for certain 
``next-to-extremal'' values of the charges, and this has subsequently 
been confirmed by an AdS calculation \cite{erdPer}. Related works
investigate the existence of other non-renormalised SCFT correlators
\cite{arufrov2} and the special structure of ``near-extremal''
correlation functions, where quantum corrections do occur \cite{defpv}. 

In this article we extend the results of \cite{ourExt} to $n$ 
points for both extremal and next-to-extremal correlators. These 
results hold for arbitrary $N=2$ superconformal gauge theories. 

In \cite{ourExt} we began by studying four-point correlators 
directly using the ideas mentioned in the first paragraph above, 
namely H-analyticity, G-analyticity and superconformal symmetry. 
However, it is in fact quicker to employ the reduction formula 
first discussed in the SCFT context in \cite{ken1}. This formula 
relates the derivative of an $n$-point correlator with respect to 
the coupling to an $(n+1)$-point correlator with one integrated insertion 
of the on-shell action; it was used in \cite{ehw} to give a proof 
of the non-renormalisation theorem for two- and three-point 
functions for arbitrary operators corresponding to Kaluza-Klein 
multiplets in AdS in $N=4$ SYM theory. Strictly speaking, this 
proof, and indeed the discussions in \cite{ourExt} and in the 
current paper on extremal correlators, depends on the assumption 
that no contact terms need to be taken into account in the 
integrated insertion despite the fact that there is an integration 
which could involve coincident points. 
In principle rather special terms of this type could affect the derivation of
these results. 
Some contact  terms have been observed in perturbation theory \cite{hsw,grkp,petske}, 
but these were of a type that do not affect the results we are interested in establishing. A search using superconformal symmetry failed to find any dangerous ones \cite{hssw} and there is no evidence that they exist from perturbation theory calculations 
\cite{toappear,3loop}. There is thus no indication so far that contact terms will affect the validity of the non-renormalisation theorems, despite fears to the contrary \cite{petske}.

In $N=2$ SYM theory the reduction formula, when applied to a 
correlation function of $n$ hypermultiplet composites, relates its 
derivative with respect to the coupling to an $(n+1)$-point 
correlator with an insertion of the chiral operator $\tr(W)^2$, 
where $W$ is the $N=2$ field strength superfield. The idea is then 
to show that this mixed $(n+1)$-point function vanishes when the 
charges of the original correlator are either extremal, which 
means that the charge of the first operator is equal to the sum of 
all of the others, or next-to-extremal, in which case the charge of 
the first operator equals the sum of all the others minus two.
{}From this we conclude that the derivative of the $n$-point 
correlator with respect to the coupling vanishes, and so there can 
be no corrections to the lowest-order free-field expression. For 
the extremal case this expression is unique while there are 
several possibilities for the next-to-extremal case. 

We shall begin with a discussion of $N=2$ theories in the harmonic 
superspace formalism of GIKOS \cite{hh, fradkin}. This is followed 
by a discussion of the same topic in coordinate form, using the 
formalism of \cite{hhh}.

\section{Extremal and next-to-extremal $n$-point correlators in $N=2$ 
harmonic superspace} 

We shall be interested in gauge-invariant operators constructed 
from the hypermultiplets of the theory and the Yang-Mills field 
strength multiplet. Each hypermultiplet has charge 1 under the 
$U(1)$ isotropy group of $SU(2)$ which defines the sphere which is 
adjoined to ordinary superspace to form harmonic superspace. The 
hypermultiplet composites are analytic with respect to the sphere 
(H-analytic) and with respect to half of the spinorial covariant 
derivatives of superspace (G-analytic). We shall denote an 
$n$-point function of such operators by $<p_1\ldots p_n>$. The 
Yang-Mills field strength superfield $W$ is chiral and we shall 
use only the operator $\tr(W)^2$ which occurs in the Yang-Mills 
part of the action. We shall denote an $(n+1)$-point function with 
$n$ hypermultiplets and one insertion of $\tr(W)^2$ by 
$<0p_1\ldots p_n>$. In the extremal case the charges satisfy 
$p_1=p_2 +p_3 + \ldots p_n$, while in the next-to-extremal case we 
have $p_1=p_2 +p_3 + \ldots p_n - 2$. The reduction formula (see 
\cite{ourExt} for a derivation in $N=2$ superspace) states that 

\be
{\del\over\del\t}<p_1\ldots p_n>\sim \int\ d^4x_0\,d^4\th_0 
<0p_1\ldots p_n> \ee 

where $(x_0,\th_0)$ are the coordinates at the chiral point and 
$\t$ is the usual complex Yang-Mills coupling constant. The points 
on the LHS are taken to be non-coincident and, as we remarked in 
the introduction, it is assumed that there are no contact 
contributions to the integral on the RHS.

The function $<0p_1\ldots p_n>$ is chiral at point 0 and 
G-analytic at points $1,\ldots,n$, it has the corresponding 
superconformal properties and  carries positive charges 
$p_1,p_2,\ldots, p_n$ at points $1,2,\ldots,n$; it is also 
H-analytic, 
\begin{equation}\label{8}
   D^{++}_r\langle 0 
p_1p_2\ldots p_n\rangle = 0\;, \quad r=1,\ldots,n \quad \mbox{if 
point 0 $\neq\ldots\neq$ point $n$}\;. 
\end{equation}  
In addition, it has the $R$-weight of 4 left-handed $\theta$'s, as 
required by the chiral superspace integral at point 0 in the 
reduction formula. 

We shall prove the following result: 

A sufficient condition for the vanishing of $<0p_1\ldots p_n>$ is  
\begin{equation}\label{21}
 p_1 >  p_2 + \ldots + p_n - 4\;. 
\end{equation}

This means that the extremal and next-to-extremal  correlators are 
ruled out. Note that we shall not attempt to find out the most 
general conditions for the vanishing  of $<0p_1\ldots p_n>$. This 
would require a detailed study of the structure of the nilpotent 
covariants which is difficult to carry out; it is not clear that this would lead
to any new results of a similar kind.

To prove this, consider first the leading term in  $<0p_1\ldots 
p_n>$. To construct it we need a set of odd variables which are 
invariant under $Q$ supersymmetry and under the shift-like part of 
$S$ supersymmetry \cite{hssw}. $Q$ supersymmetry obviously 
suggests to use the combinations 
\begin{equation}\label{11}
  \theta^{\alpha}_{0r} = \theta^{i\alpha}_0u^+_{ri} - \theta^{+\alpha}_r\;, 
\quad \delta_Q \theta^{\alpha}_{0r} = 0\;,  \quad r=1,\ldots,n\;. 
\end{equation}
Then we can form the following two cyclic combinations of three 
$\theta^{\alpha}_{0r}$: 
\begin{equation}\label{12}
 (\xi_{12r})_{\dot\alpha} = (12)\rho_{r\dot\alpha} + 
(2r)\rho_{1\dot\alpha} + (r1)\rho_{2\dot\alpha}\;, \quad 
r=3,4,\ldots,n  
\end{equation}
where 
\begin{equation}\label{12'}
  \rho_{r\dot\alpha} = x^{-2}_{0r}(x_{0r}\theta_{0r})_{\dot\alpha}
\end{equation}
and $x_{0r}\equiv x_{L0}-x_{Ar}$ are translation-invariant and 
$(rs)\equiv u^{+i}_r u^+_{si}$ are $SU(2)$-invariant combinations 
of the space-time and harmonic coordinates, correspondingly. It is 
now easy to check that $\xi_{12r}$ are completely shift-invariant, 
i.e., 
\begin{equation}\label{13}
 \delta_{Q+S}\xi_{12r} = O(\theta^2)\;.
\end{equation}
Here one makes use of the harmonic cyclic identity 
\begin{equation}\label{131}
  (rs)t_i + (st)r_i + (tr)s_i = 0\;.
\end{equation}

Note the choice we have made: point 1 which carries the highest 
charge according to (\ref{21}) is one of the two common points in 
the set of independent variables (\ref{12}). 

Now, taking into account the required $R$-weight of the correlator  
$<0p_1\ldots p_n>$, we can write down its leading term  in the 
following form: \bea   &&\langle 0 p_1\ldots p_n\rangle  =  
\nonumber\\ 
&&\sum^n_{a,b,c,d=3}\xi_{12a}\xi_{12b}\xi_{12c}\xi_{12d} 
F^{p_1-4\vert p_2-4\vert\ldots p_a-1\ldots p_b-1\ldots p_c-1\ldots 
p_d-1\ldots\vert p_n}_{abcd}(x,u) + O(\theta^5\bar\theta) 
\label{14} \eea where the Lorentz indices have been suppressed. 
The coefficient function $F$ depends on the space-time and 
harmonic variables and carries $U(1)$ charges to match those of 
the correlator and of the nilpotent prefactor. We want to study 
the consequences of the H-analyticity condition (\ref{8}). It 
turns out that in order to prove (\ref{21}) it will be sufficient 
to look at the terms not containing $\theta_0$, $\theta^+_1$ and 
$\theta^+_2$. In this case the variables (\ref{12}) become very 
simple: 
\begin{equation}\label{1}
  \xi_{12a} \sim (12)\theta^+_a
\end{equation}
(the space-time factor is of no importance). Consequently, eq. 
(\ref{14}) is reduced to 
\begin{equation}\label{2}
   \langle 0 
p_1\ldots p_n\rangle \sim 
\sum^n_{a,b,c,d=3}\theta^+_a\theta^+_b\theta^+_c\theta^+_d 
f^{p_1\vert p_2\vert\ldots p_a-1\ldots p_b-1\ldots p_c-1\ldots 
p_d-1\ldots\vert p_n}_{abcd} 
\end{equation}
where 
\begin{equation}\label{3}
  f_{abcd}\equiv (12)^4 F_{abcd}\;. 
\end{equation}
Note that each of the coefficients $f_{abcd}$ is associated to a 
single and unique nilpotent structure 
$\theta^+_a\theta^+_b\theta^+_c\theta^+_d $. This means that we 
have to impose the  H-analyticity condition (\ref{8}) on each of 
the $f_{abcd}$'s independently. We recall that H-analyticity for a 
harmonic function of positive charge simply means that it is a 
polynomial in the harmonics of degree equal to the charge. Our 
functions (\ref{3}) have to be $SU(2)$ invariant polynomials in 
the $n$ sets of harmonics. Then it becomes clear that, taking into 
account the restriction (\ref{21}) on the charges, it is not 
possible to match $p_1$ copies of $u^+_{1i}$ with the remaining 
harmonics $u^+$ at points $2,3,\ldots,n$, so all the coefficients 
must vanish. 

We now turn to nilpotent invariants of subleading order, i.e. 
those involving $\bar\theta^+$'s in their expansion. The simplest 
one of them has 5 $\theta$'s and one $\bar\theta$. In the 
left-handed sector we can still use the $Q-$ and $S-$supersymmetry 
shift-invariant variables $\xi_{12a}$. In principle, in the 
right-handed sector we should employ the analogous  
shift-invariant variables made out of four $\bar\theta^+$'s (see 
\cite{ehw}). However, they are very complicated and have a rather 
non-trivial harmonic dependence. Fortunately, we do not need them 
in the present context. It is sufficient to use variables which 
are only $Q$ supersymmetric: 
\begin{equation}\label{4}
\bar\theta_{12a} = (12)\bar\theta_a + (2a)\bar\theta_1 + 
(a1)\bar\theta_2 \;. 
\end{equation}
This means that the term of order $5+1$ will contain more 
coefficients compared to the true $Q-$ and $S-$covariant one, but 
our condition (\ref{21}) turns out sufficient to eliminate all of 
them. Indeed, the general form of such a term is 
\begin{equation}\label{5}
\sum^n_{a,\ldots,f=3}\xi_{12a}\xi_{12b}\xi_{12c}\xi_{12d} 
\xi_{12e}\bar\theta_{12f} F^{p_1-6\vert p_2-6\vert\ldots 
p_a-1\ldots p_f-1\ldots\vert p_n}_{abcdef}\;. 
\end{equation}
Once more, we are only interested in terms  not containing 
$\theta_0$, $\theta^+_1$ and $\theta^+_2$, so (\ref{5}) can be 
reduced to 
\begin{equation}\label{6}
  \sum^n_{a,\ldots,f=3}\theta^+_a\ldots \theta^+_e\bar\theta^+_{f}
f^{p_1\vert p_2\vert\ldots p_a-1\ldots p_f-1\ldots\vert 
p_n}_{abcdef} 
\end{equation} 
where 
\begin{equation}\label{77}
  f_{abcdef} \equiv (12)^6 F_{abcdef} \;.
\end{equation}
It then becomes clear that the following restriction on the 
charges: 
\begin{equation}\label{100}
 p_1 >  p_2 + \ldots + p_n - 6
\end{equation}
will be sufficient to kill all such coefficients. In fact, this 
condition clearly follows from the extremal or next-to-extremal 
constraints on the charges, so that the result holds to this 
order. The generalisation to higher-order nilpotents is obvious 
and so the result is established. 

\section{The coordinate approach}

The above result can also be derived in the coordinate formalism, 
as we shall now sketch. We are interested in the correlator 

\be
G=<0p_1\ldots p_n> \ee 

where we have one insertion of the chiral operator $\tr(W^2)$ at 
point $0$ and $n$ hypermultiplets of with charges $p_r$ at the 
other $n$ points, $1,\ldots n$. The Ward identity reads 

\be
\left((V_0 +2\D_0) + \sum_{r}(V_r + p_r\D_r)\right) G=0 \ee 

where $V_0$ and $V_r$ are the superconformal Killing vectors in 
chiral superspace and analytic superspace respectively and $\D_0$ 
and $\D_r$ are the corresponding weight functions. These Ward 
Identities have been written out in detail elsewhere and we refer 
the reader to the literature for the details \cite{hw1}. In the present 
context we shall only need to use supersymmetries, dilations, 
R-symmetry and internal ($SL(2)$) transformations (for simplicity, 
we work in complex spacetime). The main difference with the 
harmonic formalism is that the internal transformations now appear 
as (holomorphic) conformal transformations of $\bbC P^1$. Thus we 
have internal translations, dilations and ``conformal boosts''. 
The internal dilations correspond to the $U(1)$ transformations in 
the harmonic formalism. 

Initially, the space on which we are working has coordinates 
$(x_0^{\a\adt},\th_0^{\a 1}\equiv\th^{\a},\th_0^{\a 
2}\equiv\vf^{\a})$ and $(\,x_r^{\a\adt},\l_r^{\a},\p_r^{\adt},y_r)$ 
where $x_0$ and $x_r$ are chiral and analytic $x$'s respectively 
and the $y_r$'s are local coordinates on the $n$ copies of $\bbC 
P^1$. 

For translations, internal translations and Q-supersymmetries the 
weight functions vanish. These transformations are 

\bea \d x_0^{\a\adt}&=&B^{\a\adt}-\bar\e_1^{\adt}\th^{\a} 
-\bar\e_2^{\adt}\vf^{\a} \nno \\ \d x_r^{\a\adt}&=&B^{\a\adt} -\e^{\a 
1}\p_r^{\adt} - \bar\e_2^{\adt} \l_r^{\a} \nno \\ \d y_r&=&B \nno \\ \d 
\th^{\a}&=&\e^{\a 1}\\ \d \vf^{\a}&=&\e^{\a 2} +B\th^{\a} \nno \\ \d 
\l_r^{\a}&=&\e^{\a 2}-\e^{\a 1}y_r \nno \\ \d 
\p_r^{\adt}&=&\bar\e_1^{\adt}+\bar\e_2^{\adt}y_r \nno \eea 

where the $\e$ parameters are the supersymmetry parameters and the 
$B$'s are the translational parameters. It is easy to solve the 
Ward identities for these transformations to eliminate four 
spinorial coordinates, one $x$ and one $y$. Doing this, one finds 
that $G$ can be taken to be a function of the following variables: 
$(x_{0r}^{\a\adt}, \z_r^{\a}, \p_{12r}^{\adt},y_{1r})$. Here, we 
use $y_{rs}=y_r-y_s$ and similarly for other coordinate 
differences, although the $x$ difference variables require a 
nilpotent correction. Thus we have 

\bea x_{0r}^{\a\adt}&=& 
x_0^{\a\adt}-x_r^{\a\adt}-\th^{\a}\p_r^{\adt}+{1\over(n-1)}\sum_{s,s\neq 
r}{\z_r^{\a}\p_{rs}^{\adt}\over y_{rs}} \nno \\ 
\z_r^{\a}&=&y_r\th^{\a}+\l_r^{\a}-\vf^{\a} \\ 
\p_{12r}&=&y_{12}\p_r^{\adt}+y_{r1}\p_2^{\adt} 
+y_{2r}\p_1^{\adt} \nno \\ y_{1r}&=&y_1-y_r \nno \eea 

Note that we have $n$ $x$'s and undotted spinor coordinates, 
$(n-2)$ dotted spinor coordinates and $(n-1)$ $y$ coordinates 
left. We should remark that the choice of correction terms for the 
$x$ difference variables is not unique but this is not an issue 
which will be relevant for the ensuing discussion. The weights of 
these coordinates under (dilations, internal dilations, R) are as 
follows 

\bea x&:&(1,0,0) \nno \\ y&:&(0,1,0) \\ \z&:&({1\over2},{3\over2},1) \nno \\ 
\p&:&({1\over2},{3\over2},-1) \nno \eea 

For R symmetry $\D_0=-2R$, so that G has R-weight four. 
Schematically it must depend on the odd variables in the following 
way 

\be
G\sim \z^4(1 + {\rm power\ series\ in}\ (\z\p)) \ee 

The idea is then to carry out the analysis order by order in odd 
variables. Since we shall only proceed to the second order when 
the first order vanishes this means that we can ignore the 
correction terms in $x_{0r}$ and that we can simplify the 
remaining S-supersymmetry transformations considerably. The only 
non-trivial simplified S-supersymmetry transformation with dotted 
parameters is 

\be
\d \z_r^{\a}=(-\bar\h_{\bdt}^2 + 
y_r\bar\h_{\bdt}^1)x_{or}^{\a\bdt} \ee 

It is easy to find variables invariant under these transformations 
since they are essentially translations. These invariant variables 
are 

\be
\xi_{12r}^{\adt}=y_{12}(x_{0r}^{-1})_{\adt\a}\z_r^{\a}+ 
y_{r1}(x_{02}^{-1})_{\adt\a}\z_2^{\a}+ 
y_{2r}(x_{01}^{-1})_{\adt\a}\z_1^{\a} \ee 

The $\xi$'s have weights $(-1/2,3/2,1)$. Note that, in this 
approximation, all the other coordinates are invariant, so we may 
now work with the set $(x_{0r},\xi_{12r},\p_{12r},y_{1r})$. The 
leading term in $G$, $G_0$, say,  can be written as 

\be
G_0=\sum_{abcd}\xi_{12a}\xi_{12b}\xi_{12c}\xi_{12d}F^{abcd}(x,y) 
\ee 

where the Lorentz indices have been suppressed. To complete the 
analysis we need only consider the remaining internal symmetry 
transformations; we can forget about $x$ altogether. Internal 
dilational symmetry implies that $F$ has weight $(\sum_{r=1}^n 
p_r)-12$. (The chiral function $\D_0=0$ for this while 
$\D_r=-1/2$). The internal ``conformal boosts'' act in the 
following way 

\bea \d y_{rs}&=&C(y_r+y_s)y_{rs} \nno \\ \d \xi_{12r}&=&C(y_1+y_2+y_r) 
\xi_{12r} \\ \d \pi_{12r}&=&C(y_1+y_2+y_r) \pi_{12r} \nno \eea 

where $C$ is the parameter. If we define an operator $D$ by 

\be
D=\sum \left( {\d y\over C} {\del\over \del y} + {\d \xi\over 
C}{\del\over\del\xi}\ +{\d\p\over C}{\del\over\del\p}\right) \ee 

where the sum is over these coordinates with the variations given 
in the previous equation, then the corresponding Ward identity 
implies that 

\be
D G=(\sum_{r=1}^n p_r y_r)G \la{DG} \ee 

since $\D_0=0$ and $\D_r=Cy_r$, up to nilpotent terms arising from 
the variation of the $x$'s. The same equation holds (exactly) for 
$G_0$. It makes it clear that we can identify the internal 
dilation charges in the coordinate approach with the $U(1)$ 
charges in the harmonic formalism. 

{}From \eq{DG} we find

\be
\sum_{abcd} \xi_{abcd}\left(D F^{abcd} +(4y_1 + 4y_2 + y_a +\ldots 
y_d)F^{abcd}\right)= (\sum_{r=1}^n p_r 
y_r)\sum_{abcd}\xi_{abcd}F^{abcd} \ee 

where we have used the abbreviation 
$\xi_{abcd}\equiv\xi_{12a}\xi_{12b}\xi_{12c}\xi_{12d}$. Now 
$\xi_{abcd}$ includes the term $(y_{12})^4\l_a\l_b\l_c\l_d$, and 
the only place such a term appears in the sum for given values of 
$a,b,c,d$ is as a term in the corresponding $\xi_{abcd}$. {}From 
this we conclude that, for all choices of $a,b,c,d$, 

\be
D F^{abcd}=(\sum_{r=1}^n p'_r y_r) F^{abcd} \ee 

where 

\be
p'_r=\cases{p_r-4, &\quad $r=1,2$ \cr p_r,&\quad $r\neq 
1,2,a,b,c,d$ \cr p_r-1,&\quad $r=a,b,c,d $} \ee 

In other words,  any such $F^{abcd}$ is a conformal function  with 
the above weights. Now $G$ is analytic in the internal 
coordinates. Analyticity of the term in $G_0$ in which $F^{abcd}$ 
is multiplied by $(y_{12})^4\l_a\l_b\l_c\l_d$ implies that 
$F^{abcd}$ cannot have any singularities in $y_{rs}$ for $r<s, 
s\geq 3$.  It can therefore be written as a sum of terms of the 
form 

\be
F^{abcd}=(y_{12})^{q_2}\ldots (y_{1n})^{q_n} \tilde F^{abcd} \ee 

where $\sum_{r=2}^n q_r=p_1-4$ and where each $\tilde F^{abcd}$ 
depends only on  the coordinates $y_{rs}$ with $2\leq r<s\leq n$. 
$\tilde F^{abcd}$ has charges $\tilde p_r, $ where 

\be
\tilde p_r=\cases{0,&\quad $r=1$ \cr p_r-4-q_r,&\quad $r=2$ \cr 
p_r-q_r,&\quad $r\neq 1,2,a,b,c,d$ \cr p_r-q_r-1,&\quad $r=a,b,c,d 
$} \ee 

By analyticity, $\tilde F^{abcd}$ must be a homogeneous polynomial 
in $y_{rs}, 2\leq r<s\leq n$, and so $\tilde p_r\geq 0, r\neq 1$. 
On summing these constraints we find 

\be
p_1-4=\sum_{r=2}^n q_r\leq \sum_{r=2}^n p_r -8 \ee 

Hence there is no solution if $p_1>\sum_{r=2}^n p_r-4$, which 
covers the extremal and next-to-extremal cases. 

Given that the lowest order term vanishes, at the next order we 
can write 

\be
G_1=\sum_{abcdef}\xi_{12a}\xi_{12b}\xi_{12c}\xi_{12d}\xi_{12e}\pi_{12f} 
F^{abcdef} \ee 

Now since $\p_{12r}$ transforms in the same way as $\xi_{12r}$ 
under conformal boosts it follows that we can repeat the above 
argument straightforwardly. The only difference is that the bound 
is stronger; there is no solution if $p_1>\sum_{r=2}^n p_r-6$. 
Clearly the argument extends to all orders in this manner.

Finally we remark that although the above results for $N=2$ cover a wider class of theories than $N=4$ they are in another sense more restricted. This is because the $N=2$ hypermultiplet operators have maximum spin $1$ while the single-trace analytic oper
ators in $N=4$ have maximum spin $2$. It should in principle be possible to carry out a similar sort of analysis to the one given here for $N=2$ directly in $N=4$ harmonic superspace, but it would be considerably more complicated due to the fact that the 
internal symmetry group is significantly larger. On the other hand it is not difficult to show that the leading terms of correlators of such operators are completely determined, by group theory, from $N=2$ hypermultiplet ``component'' correlators in both 
the extremal and next-to-extremal cases. Thus the leading terms of these correlators are trivial and it seems very likely that this result will extend to all orders in an expansion with respect to the third and fourth spinorial coordinates.

\vspace{20pt} {\bf Acknowledgements:} 
This work 
was supported in part by the British-French scientific programme 
Alliance (project 98074), by the EU network on Integrability, 
non-perturbative effects and symmetry in quantum field theory 
(FMRX-CT96-0012), by the grant INTAS-96-0308 and  by PPARC through SPG 613.

\end{document}